\documentclass{article}
\usepackage{graphicx}
\usepackage{epsfig}
\def \BE {\begin{equation}}
\def \EE {\end{equation}}
\def \BEAH {\begin{eqnarray*}}
\def \EEAH {\end{eqnarray*}}
\def \BEA {\begin{eqnarray}}
\def \EEA {\end{eqnarray}}
\def \BDM {\begin{displaymath}}
\def \EDM {\end{displaymath}}
\def \pul {{{\footnotesize{\frac{1}{2}}}}}
\def \eqv4d {\stackrel{4D}{\Longleftrightarrow }}

\def \nn {\nonumber}

\def \bl {\mbox{\boldmath{$\ell$}}}
\def \hbl {\mbox{\boldmath{$\hat \ell$}}}
\def \bn {\mbox{\boldmath{$n$}}}
\def \hbn {\mbox{\boldmath{$\hat n$}}}
\def \bm #1 {\mbox{\boldmath{$m^{(#1)}$}}}
\def \hbm #1 {\mbox{\boldmath{$\hat m^{(#1)}$}}}

\begin{document}

\vspace*{3cm}

\begin{center}
{\bf \Large WANDs of the Black Ring}\\
\vspace*{2cm}
{\bf V. Pravda}\footnote{E-mail: {\tt
pravda@math.cas.cz}}
{\bf and A. Pravdov\'a}\footnote{E-mail: {\tt pravdova@math.cas.cz}}

\vspace*{1cm}

{\it Mathematical Institute, 
Academy of Sciences, \v Zitn\' a 25, 115 67 Prague 1, Czech Republic}\\[1cm]

\end{center}

\begin{abstract}
Necessary conditions for various algebraic types of the Weyl tensor are determined.
These conditions are then used to find Weyl aligned null directions for the
black ring solution. It is shown that the black ring solution is algebraically special, of type ${\rm I}_i$,
while locally  on the horizon the type is II. One exceptional subclass -- the Myers-Perry 
solution -- is of type D.
\end{abstract}
\section{Introduction}

Recently a classification of tensors on Lorentzian manifolds of arbitrary dimensions was introduced
\cite{Algclass}. When applied to the Weyl tensor \cite{Algclass,Weylletter}, in four dimensions it reduces to the
well known Petrov classification, in higher dimensions it leads to a similar, dimensionally independent 
classification scheme (see Table 1). This classification is based on the existence of preferred null directions
-- Weyl aligned null directions (WANDs) and corresponding principal null congruences in a given spacetime (see Section \ref{classif} for details). 
In 4 dimensions WANDs correspond to principal null directions of the Weyl tensor. 

The Petrov classification in four dimensions was very useful for generating algebraically special exact
solutions as well as for physical interpretation of spacetimes. Ultimately, one would like to use the higher dimensional
classification for similar purposes as in classical relativity, 
however, while in higher dimensions some
properties related to the classification remain the same as in four dimensions,  other differ.

For example, it was proven  that for all vacuum spacetimes of types III and N in arbitrary dimension the principal
null congruence is geodesic as in four dimensions \cite{Bianchi}. This congruence in four dimensions is in addition shear-free but
in higher dimensions shear may not vanish (see \cite{Bianchi} for additional comments on the Goldberg-Sachs theorem in higher dimensions).
Also all spacetimes with vanishing curvature invariants are necessarily of types III and N in arbitrary dimension, including four \cite{HDVSI}.

Black holes are another interesting example. 
First let us point out that the Kerr solution as well as the Myers-Perry solution in five
dimensions are of type D with geodesic principal null congruences \cite{Bianchi,Frolov}. More general black holes in four dimensions
can be algebraically general, however, if the horizon is isolated, the spacetime is of type II locally on the horizon \cite{IsolHoriz}.
Recently, it was proven  that locally, on the isolated horizon, the algebraic type is  II also in higher dimensions \cite{IsolHD}. 

A vacuum, asymptotically flat, stationary black hole solution with a horizon of topology $S^1 \times S^2$ - so called black ring -  
was discovered recently in \cite{ER-PRL}. There exist black holes of spherical topology and black rings with the same values of the conserved quantities
$M$ and $J$ and thus black holes in higher dimensions are not uniquely characterized by their mass and angular momenta and 
uniqueness theorems cannot be (straightforwardly) generalized for higher dimensions. 

In this paper it is shown that  while the Myers-Perry solution representing a spherical black hole 
is of type D,  black rings, while still algebraically special, belong to the more general class ${\rm I}_{i}$.
In Section \ref{classif} the classification of the Weyl tensor in higher dimensions is overviewed  and necessary
conditions for various classes, which significantly simplify search for WANDs, are determined.
In Section \ref{secring} we briefly overview the (neutral) black ring solution and analyze some of its properties.
In Section \ref{secbrclass} we  classify static and stationary black rings.
In four dimensions it is possible to determine  algorithmically an algebraic type of a solution using 
various invariants and covariants of the metric
(see e.g. Chapter 9 in \cite{Stephani}).  Since such a method  is not developed in higher dimensions,
it is necessary to 
 find  WANDs explicitly. Components of the Weyl tensor for the black ring are quite complicated and thus
 corresponding necessary conditions are also complicated even with the use of Maple (however, much simpler than
alignment equations). We thus limit ourselves to presenting
only their solutions and indicating the procedure how to obtain them.
We also
explicitly show that locally on the horizon the Weyl tensor is of type II.

\section{Algebraic classification of the Weyl tensor in higher dimensions}
\label{classif}
In this section,  algebraic classification of the Weyl tensor in higher dimensions developed in  
\cite{Algclass,Weylletter} is briefly summarized  and at the end  necessary
conditions for various algebraic types are introduced. 

We will work in the  frame
\BDM
\bm{0} 
= \bn, \ \bm{1} 
=\bl, \ \bm{i} 
 , \ \ \ i,j,k = 2 \dots D-1,
\EDM
with two null vectors   $\bn,\ \bl$  
\BDM
\ell^a \ell_a= n^a n_a = 0,\ \   \ell^a n_a = 1, \ \  a = 0  \dots D-1,
\EDM
 and $D-2$ spacelike vectors $\bm{i} $ 
\BEAH
 m^{(i)a}m^{(j)}_a=\delta_{ij},\ \ m^{(i)a}\ell_a=0=m^{(i)a}n_a,  \ \ \ \ \ \ i,j,k = 2 \dots D-1, 
\EEAH
where 
the {metric} has the form
\BDM
g_{a b} = 2 \ell_{(a}n_{b)} + \delta_{ij} m^{(i)}_a m^{(j)}_b  . \label{metric} 
\EDM

The group of ortochronous Lorentz transformations is generated by 
{null rotations}
\BE
\hbl =  \bl +z_i {\bm{i} } -\frac{1}{2} z^i z_i\, \bn , \ \    
\hbn =  \bn, \ \ 
    \hbm{i} =  \bm{i} - z_i \bn 
    \label{nullrot}
\EE
 { spins}  and {boosts}, respectively,
\BDM
\hbl =  \bl, \ \ \hbn = \bn, \ \hbm{i} =  X^{i}_{\ j} \bm{j} ;\ \ \ \ \ \ \ \ \ \ 
\hbl = \lambda \bl, \ \  \hbn = \lambda^{-1} \bn, \ \ \hbm{i} = \bm{i} .
\EDM
  A quantity   $q$   has  a {\it boost weight}   $b$   if it transforms under a boost according to
$\hat q = \lambda^b q  $. 
{\it Boost order} of a tensor ${\mathbf T}$ is defined  as the maximum boost weight of its frame components
and it can be shown that
%
it depends only on the choice of a null direction  $\bl$ (see Proposition III.2 in \cite{Algclass}).
For a given tensor  ${\mathbf T}$,   $b_{\rm{max}}$    denotes the maximum value 
of   $b({\mathbf k})$   taken over all null vectors   ${\mathbf k}$ .
Then a null vector   ${\mathbf k}$   is {\it aligned}  with the tensor   ${\mathbf T}$  
whenever    $b({\mathbf k}) < b_{\rm{max}}$   and  the integer   $b_{\rm{max}}-b({\mathbf k})-1$
  is called {\it order of alignment}.
The classification of tensors \cite{Algclass} is based on the existence of such aligned null vectors of various orders.
Namely, for the Weyl tensor  the {\it primary alignment type} is G if there are no null vectors aligned with the Weyl tensor
and the primary alignment type is 1, 2, 3, 4  if the maximally aligned null vector has
order of alignment 0, 1, 2, 3, respectively.
Once  $\bl$ is fixed as 
an aligned null vector of maximal order of alignment, one  can search for $\bn$ with 
maximal order of alignment subject to the constraint $\bn \cdot \bl=1$ and 
similarly define {\it secondary alignment types}.

Let us introduce the operation \{ \} which allows us to construct a basis in the space of Weyl-like tensors by
\BE
w_{\{a} x_b y_c z_{d\}}  \equiv \frac{1}{2} (w_{[a} x_{b]} y_{[c} z_{d]}+ w_{[c} x_{d]} y_{[a} z_{b]}).
\label{zavorka}
\EE
Now we can decompose the Weyl tensor in its frame components and
sort them 
by their boost weight (see \cite{Algclass}): 
\BEAH
  C_{abcd} &=& 
  \overbrace{
    4 C_{0i0j}\, n^{}_{\{a} m^{(i)}_{\, b}  n^{}_{c}  m^{(j)}_{\, d\: \}}}^2  
  +\overbrace{
    8C_{010i}\, n^{}_{\{a} \ell^{}_b n^{}_c m^{(i)}_{\, d\: \}} +
    4C_{0ijk}\, n^{}_{\{a} m^{(i)}_{\, b} m^{(j)}_{\, c} m^{(k)}_{\, d\: \}}}^1  
  \nonumber \\&+& \overbrace{
      4 C_{0101}\, \, n^{}_{\{a} \ell^{}_{ b} n^{}_{ c} \ell^{}_{\, d\: \}} 
\;  + \;  C_{01ij}\, \, n^{}_{\{a} \ell^{}_{ b} m^{(i)}_{\, c} m^{(j)}_{\, d\: \}} 
      +8 C_{0i1j}\, \, n^{}_{\{a} m^{(i)}_{\, b} \ell^{}_{c} m^{(j)}_{\, d\: \}}
   +  C_{ijkl}\, \, m^{(i)}_{\{a} m^{(j)}_{\, b} m^{(k)}_{\, c} m^{(l)}_{\, d\: \}}
    }^0
  \label{eq:rscalars}\\ 
   &+& \overbrace{
    8 C_{101i}\, \ell^{}_{\{a} n^{}_b \ell^{}_c m^{(i)}_{\, d\: \}} +
    4 C_{1ijk}\, \ell^{}_{\{a} m^{(i)}_{\, b} m^{(j)}_{\, c} m^{(k)}_{\, d\: \}}}^{-1} 
  + \overbrace{
      4 C_{1i1j}\, \ell^{}_{\{a} m^{(i)}_{\, b}  \ell^{}_{c}  m^{(j)}_{\, d\: \}}}^{-2} .
\EEAH
Additional constraints follow from symmetries of the Weyl tensor
\BEA 
  &&C_{0[i|0|j]} = 0,\nn\\
  \nonumber
  &&C_{0i(jk)}=C_{0\{ijk\}} = 0,\\
  &&C_{ijkl} = C_{\{ijkl\}},\quad C_{i\{jkl\}}=0,\quad 
  C_{01ij}= 2 C_{0[i|1|j]}, \label{eq:rcomps}\\
  \nonumber
  &&C_{1i(jk)}=C_{1\{ijk\}} = 0, \\
  \nonumber
  &&C_{1[i|1|j]} = 0
\EEA
and from its tracelessness 
\BEA
  \label{eq:rfree}
  && C_{0i0i} =  C_{1i1i} =0,\\
  \nonumber
  &&C_{010i} = C_{0jij},\quad C_{101i} = C_{1jij},\\
  \nonumber
   && 2C_{0i1j}=C_{01ij}- C_{ikjk},\quad 
  C_{0101} = -\pul C_{ijij}.
\EEA 

Now the Weyl tensor is of type I if there exist such $z_i$ that we can 
set all components of boost weight 2, $C_{0i0j}$,  to zero using the transformation (\ref{nullrot}). It is of type II if we can set boost order 2 and
boost order 1 components to zero and so on. Resulting polynomial equations for $z_i$ depend on the original choice 
of the frame $\bl,\ \bn, \ \bm{i} $ and can be considerably simplified if the frame is chosen appropriately. However,
there is no general method how to make such a choice. For this reason we investigate another method, which involves
only  the vector $\bl$ instead of the frame $\bl,\ \bn, \ \bm{i} $.

In four dimensions, the  following equivalences for principal null directions hold  (see e.g. \cite{Stephani})
\BEA
\begin{tabular}{rcl}
$\ell^b \ell^c \ell_{[e}C_{a]bc[d}\ell_{f]}=0$ & $\eqv4d $ & $\ell$ is PND, at most Petrov type I;\\
$\ell^b \ell^c C_{abc[d}\ell_{e]}=0$& $\eqv4d $& $\ell$ is PND, at most Petrov type II;\\ 
$ \ell^c C_{abc[d}\ell_{e]}=0$ & $\eqv4d $ & $\ell$ is PND, at most Petrov type III;\\ 
$\ell^c C_{abcd}=0$ & $\eqv4d $ & $\ell$ is PND, at most Petrov type N.\\
\label{PND4D}
\end{tabular}
\EEA

By substituting a general form of the Weyl tensor of various primary types in previous equations (\ref{PND4D}) we can find that in arbitrary dimension
\BEA
\begin{tabular}{rcl}
$\ell^b \ell^c \ell_{[e}C_{a]bc[d}\ell_{f]}=0$ & $\Longleftarrow  $ & $\ell$ is WAND, at most primary type I;\\
$\ell^b \ell^c C_{abc[d}\ell_{e]}=0$& $\Longleftarrow  $& $\ell$ is WAND, at most primary  type II;\\ 
$ \ell^c C_{abc[d}\ell_{e]}=0$ & $\Longleftarrow  $ & $\ell$ is WAND, at most primary type III;\\ 
$\ell^c C_{abcd}=0$ & $\Longleftarrow  $ & $\ell$ is WAND, at most primary type N.\\
\end{tabular}
\label{HDnecessity}
\EEA
In fact for the type I equivalence holds in arbitrary dimension \cite{Algclass} but it is not so
for more special types. 
For example, it can be shown that the most general Weyl tensor satisfying $\ell^c C_{abcd}=0$
has the form 
\BE
  C_{ijkl}\, \, m^{(i)}_{\{a} m^{(j)}_{\, b} m^{(k)}_{\, c} m^{(l)}_{\, d\: \}}
   + 4 C_{1ijk}\, \ell^{}_{\{a} m^{(i)}_{\, b} m^{(j)}_{\, c} m^{(k)}_{\, d\: \}}
  + 4 C_{1i1j}\, \ell^{}_{\{a} m^{(i)}_{\, b}  \ell^{}_{c}  m^{(j)}_{\, d\: \}}.
\EE
Its quadratic curvature invariant  is
$C_{abcd} C^{abcd} = C_{ijkl} C^{ijkl} = \Sigma (C_{ijkl})^2$,
which is non-zero as long as $C_{ijkl}$ has a non-vanishing component. Note that due to symmetries of $C_{ijkl}$ it can have
a non-vanishing component only for dimensions $D \geq$ 6. Since the Weyl tensor possesses a non-vanishing invariant, it cannot be of type 
N or III and thus the equivalence in general does not hold.

Let us conclude with a table comparing the classification of the Weyl tensor in four and higher dimensions.

\begin{table}[h]
\begin{center}
\begin{tabular}{|cc|c|}
\hline
\ \ D$>$4 dimensions & & 4 dimensions \\
\hline
Petrov type & alignment type& Petrov type   \\
\hline
G     & G &      \\
I     & (1)   &  \\
${\rm I}_{i}$ & (1,1) & I \\
II    & (2)  &  \\
${\rm II}_{i}$ & (2,1) & II \\
D & (2,2) & D \\
III & (3) &  \\
${\rm III}_{i}$ & (3,1) & III \\
N & (4) & N \\
\hline
\end{tabular}
\caption{Comparison of the algebraic classification of the Weyl tensor in four and higher dimensions. Note that in four dimensions
alignment type (1) is necessarily equivalent to the type (1,1), (2) to (2,1) and (3) to (3,1) and that type G does not exist.} 
\end{center}
\end{table}

\section{Rotating black ring - overview}
\label{secring}

The rotating black ring solution was found in \cite{ER-PRL}, here  we will use a slightly different form
of the metric in coordinates  $\{ t,\ x,\ y,\ \phi,\ \psi \} $ introduced in \cite{EE-JHEP}
\BEA
  \nonumber
  {\rm d}s^2 &=& -\frac{F(x)}{F(y)} \left( {\rm d}t+
     R\sqrt{\lambda\nu} (1 + y) {\rm d} \psi\right)^2  \\
  &&
   +\frac{R^2}{(x-y)^2}
   \left[ -F(x) \left( G(y) {\rm d}\psi^2 +
   \frac{F(y)}{G(y)} {\rm d}y^2 \right)
   + F(y)^2 \left( \frac{{\rm d}x^2}{G(x)} 
   + \frac{G(x)}{F(x)} {\rm d} \phi^2\right)\right] ,
\label{ringmetric}
\EEA
where
\BE
  F(\xi) = 1 - \lambda\xi \, ,
\qquad  G(\xi) = (1 - \xi^2)(1-\nu \xi) \, .
\label{polFG}
\EE

Note that only certain regions in the
$(x,y)$ plane have signature +3. These regions can be determined
by analyzing eigenvalues
$\lambda_1 \dots \lambda_5$ of the metric
\BEAH
\lambda_2 &=& \frac{R^2 F(y)^2}{G(x)(x-y)^2 } , \\
\lambda_3 &=& -\frac{R^2 F(x) F(y)}{G(y) (x-y)^2},\\
\lambda_4 
      &=& \frac{R^2 F(y)^2 G(x)}{F(x) (x-y)^2},\\
\lambda_1 
\lambda_5 &=& -\frac{R^2 F(x)^2 G(y)}{F(y) (x-y)^2 }.
\EEAH
We do not give explicit forms of $\lambda_1$ and  $\lambda_5$ but only their product which is much simpler.
For $\lambda_1 
\lambda_5 <0$, values of $\lambda_2 
\dots \lambda_4 
$ have to be positive and we obtain
3 regions $(-1,1) \times (-\infty,-1)$, \mbox{$(-1,1) \times (1,1/\lambda)$,}  $(-1,1) \times (1/\nu,\infty)$
with signature +3. For $\lambda_1
\lambda_5 >0$, $\lambda_5 $ has to be positive and one of the eigenvalues 
$\lambda_2
\dots \lambda_4 
$ negative. We arrive at following three regions $(-1,1) \times (1/\lambda,1/\nu)$,
$(1/\lambda,1/\nu) \times (-1,1)$ and $(1/\nu,\infty) \times (-1,1)$ (see Figure 1).

The curvature invariant $R_{abcd}  R^{abcd}$ has the form
\BE
\frac{(x-y)^4 P_{\lambda \nu }(x,y)}{ R^4 (-1+\lambda x)^4 (-1+\lambda y)^6 }, \label{Kretschmann}
\EE
where the polynomial
$P_{\lambda \nu }(x,y)$ is quadratic in $y$ and a 6th degree polynomial in $x$, which does not vanish at
$ x=1/ \lambda\not=y$ or $ y=1/\lambda\not=x$.
Consequently, there are curvature singularities located at $x=1/\lambda$, $y=1/\lambda$, and 
$x=\pm \infty$. The invariant (\ref{Kretschmann}) as well as other curvature invariants vanish at $x=y$.
This indicates that the spacetime is flat there. 

Let us now summarize basic properties of various regions in the black ring solution:

Region ${\cal A}_1$: Here signs of $\lambda_1 \dots \lambda_5 $ are -++++. 
This region is asymptotically flat, $\frac{\partial}{\partial t}$ is  a timelike
Killing vector and thus spacetime is static here. Moreover, the norm of the Killing vector $\frac{\partial}{\partial t}$
approaches $-1$ at the ``flat point'' (x,y)=(-1,-1). 
This region represents an outer part of the black ring solution.
Region ${\cal A}_1$ can be smoothly connected with ${\cal A}_2$ by identifying $y=-\infty$ with $y=\infty$.  

In ${\cal A}_2$, signs of $\lambda_1 \dots \lambda_5 $ are  ++++-, 
both $\frac{\partial}{\partial t}$ and $\frac{\partial}{\partial \psi}$ are spacelike,
but there  exists a timelike Killing vector as their linear combination.  This region represents an ergosphere with a limiting surface 
of stationarity located at $y=\infty$ and a horizon at $y=1/\nu$. 

In ${\cal A}_3$, signs of $\lambda_1 \dots \lambda_5 $ are ++-++, spacetime is non-stationary and represents the region below the horizon. Curvature
singularity is located at $y=1/\lambda$.

In region ${\cal B}$, signs of $\lambda_1 \dots \lambda_5 $ are -++++. 
This region is asymptotically flat. 
In the neighbourhood of the curvature singularity
located at $y=1/\lambda$, both $\frac{\partial}{\partial t}$ and $\frac{\partial}{\partial \psi}$ are timelike and thus this region
contains closed timelike curves. In the vicinity of the ``flat point'' (x,y)=(1,1), $\frac{\partial}{\partial \psi}$ becomes spacelike and 
the norm of
$\frac{\partial}{\partial t}$ approaches -1. This region  represents a spacetime of a spinning naked singularity.

In region ${\cal C}_1$, signs of $\lambda_1 \dots \lambda_5 $ are +++-+. 
Note that $\frac{\partial}{\partial \phi}$ is timelike and thus this region
admits closed timelike curves. Curvature singularity is located at $x=\infty$. In region ${\cal C}_2$, signs of $\lambda_1 \dots \lambda_5 $ are +-+++. 
Curvature singularity is located at $x=1/\lambda$.
Regions ${\cal C}_1$ and ${\cal C}_2$ are not asymptotically
flat and their physical interpretation is unclear.
\begin{figure}
\begin{center}
\includegraphics[scale=0.7]{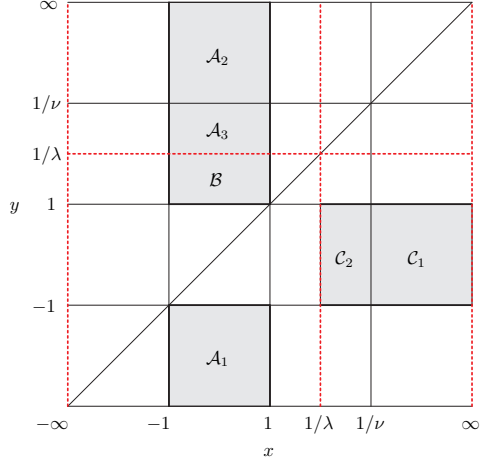}
\end{center}
\caption{Various regions in the black ring solution: regions with signature +3 are shaded; curvature singularities are located
at $x=1/\lambda$, $y=1/\lambda$, and 
$x=\pm \infty$ (dashed lines); the metric is flat at $x=y$ (see the text for details).}
\label{diagram}
\end{figure}


\section{Black ring - algebraic structure}
\label{secbrclass}

In this section we classify the black ring solution and its various special cases.
Our method is to solve the necessity conditions (\ref{HDnecessity}) and then check that these solutions indeed
represent WANDs by calculating components of the Weyl tensor in an appropriate frame.

\subsection{Myers-Perry metric is of type D}

By setting $\lambda=1$ in (\ref{ringmetric}) we obtain the Myers-Perry 
metric \cite{Myers} with a single rotation parameter.

It turns out that the second equation in (\ref{HDnecessity}) admits two independent solutions
\BEA
L_{\pm} = 
{\frac {1}{ \left( x^2-1 \right)    \left(  
-1+\nu\,y \right) }} \left(
 \frac{ \nu\,yx-y+\nu\,x+1-2\,\nu\,y  }{ x-y     }{\it R} \partial_t 
-
 {\sqrt {\nu}} \partial_{\psi} \right)
\pm 
\,\sqrt {{\frac {\nu\,x-1}{ \left( x-y \right)  \left( y-1
 \right) }}} \left( \partial_x 
+
 {\frac {{y}^{2}-1}{{x}^{2}-1}} \partial_y \right). \label{PNDMP}   
 \EEA
When we choose a frame with $\bl \sim L_{+}$ and $\bn \sim L_{-}$ all components of the Weyl tensor with boost weights 2,1,-1,-2 vanish
and the spacetime is thus of type D. These two vectors were given  in Boyer-Lindquist coordinates in \cite{Frolov} and
 further discussed 
in App. D in \cite{Bianchi}.

\subsection{Black ring  is of type II on the horizon}

The transformation
\BEA
{\rm{d}} \chi &=& {\rm{d}} \psi + {\frac{\sqrt{-F(y)}}{G(y)}} {\rm{d}} y,  \\
{\rm{d}} v &=& {\rm{d} t} - R \sqrt{\lambda \nu } (1+y) {\frac{\sqrt{-F(y)}}{G(y)}}  {\rm{d}}y
\EEA
(see \cite{ER-PRL} for a similar transformation) leads to a metric regular on the horizon $y=1/\nu$
\BEA
{\rm{d}} s^2 = && -\frac{F(x)}{F(y)} ({\rm{d}} v+\sqrt{\lambda \nu} R (1+y) {\rm{d}} \chi)^2  \nonumber \\ 
&& + \frac{R^2}{(x-y)^2} \left[  -F(x) \left(G(y)  {\rm{d}} \chi^2
 - 2 \sqrt{-F(y)}  {\rm{d}} x {\rm{d}} y \right) + F(y)^2 
\left( \frac{{\rm{d}} x^2}{G(x)}+\frac{G(x)}{F(x)} {\rm{d}} \phi^2 \right) \right].
\EEA

The second equation in (\ref{HDnecessity}) admits a solution
\BDM
L=\partial_{v}-\sqrt{\frac{\nu}{\lambda}}  \frac{1}{R(1+\nu)} \partial_x.
\EDM
One can check that boost order of the Weyl tensor in the frame with $\bl=L$ is 0   and thus the black ring is of the type II
on the horizon.

\subsection{Black ring is of type ${\rm I}_i$}

In order to solve
 the first equation in (\ref{HDnecessity})
\BE
I_{eadf} = \ell^b \ell^c \ell^{[e} C^{a]\  \ [d}_{\ \ bc} \ell^{f]}=0, \label{BRI}
\EE
we denote 
\BE
\ell^a=(\alpha, \beta, \gamma, \delta, \epsilon) \label{WAND}
\EE 
and from (\ref{BRI}) we obtain a set of fourth order polynomial equations in $\alpha \dots \epsilon$. An additional
second order equation follows from $\ell_a \ell^a=0$.
Since components of the Weyl tensor are quite complicated, 
it is not surprising that
these equations are also complicated.
However, one can  pick some of them which are relatively simple. Let us start with the static case.

\subsubsection{The static case}

The static case can be obtained  by setting  $\nu = 0$. Two particularly simple equations
\BE
I_{tx\phi\psi}= \,{\frac {3\alpha\,\beta \, \delta\, \epsilon\, \lambda\, 
\left( \lambda^2-1 \right)
\left( x-y \right) ^{3}
\,    }
{2 {{\it R}}^{2} \left( -1+\lambda\,x \right) ^{2}  \left( -1+\lambda\,y
 \right) ^{3}}}=0,
\EE

\BE
I_{xy\phi\psi}=-\,{\frac {3\beta\, \gamma\, \delta \,\epsilon\,\lambda \left( \lambda^2-1 \right) 
\left( x-y \right) ^{3} 
  }{2 {{\it R}}^{2}
 \left( -1+\lambda\,x \right) ^{2}  \left( -1+\lambda\,y
 \right) ^{3}}}=0
\EE
imply that unless $\lambda$ equals to -1,0 or 1, at least one component of the WAND (\ref{WAND})
vanishes. More detailed analysis of equations (\ref{BRI}) shows that the only non-trivial solution
is $\epsilon=0$ and 
\BE
\alpha^2=
{\frac {{{\it R}}^{2}\gamma\,  \left( -1+\lambda\,y \right) ^{2}\left[ \gamma(\lambda\,x-1)+
\beta\,(1-\lambda\,y) \right] }{
 \left( x-y \right) ^{2} \left( -1+\lambda\,x \right)  \left( y^2-1
 \right) }},
 \label{eqastatic}
\EE

\BE
{\delta }^{2}={\frac {\beta\,\left( 1-\lambda\,x \right) \left[ \gamma\,({x}^{2}-1)+\beta\,(1-{y}^{2})
 \right]   }{ \left( y^2-1 \right) 
   \left( x^2-1 \right) ^{2}}},
\label{eqdstatic}
\EE

with $\beta, \gamma $ satisfying the quadratic equation
\BEA
&& \lambda\, \left( x^2-1 \right)   \left( -1+\lambda\,x
 \right) ^{2}{\gamma}^{2}+\lambda\, \left( y^2-1 \right)  
  \left( -1+\lambda\,y \right)  \left( -1+\lambda\,x \right) {
\beta}^{2} \label{QDstatic} \\ 
&& -\lambda\,\beta\, \gamma \left\{
\lambda\,(x+y)[\lambda\,x(-1+xy)+(1-x^2)]+(x-y)^2(2-\lambda^2)+2(-1+\lambda\,x)+2xy(1-\lambda\,y)\right\}=0.
%
\nonumber
\EEA
Note that a vector as well as its arbitrary multiple  represent the same  WAND.
We can thus set one component of (\ref{WAND}) equal to a given 
number. Let us set $\gamma=\gamma_0$. Now we can solve (\ref{QDstatic}) for $\beta$. It has
in general two real roots, however, only one of them leads to positive values
of $\alpha^2$ and $\delta^2$ as given by (\ref{eqastatic}) and (\ref{eqdstatic}).
Let us denote this root by $\beta_0$ and corresponding square roots of right hand
sides of (\ref{eqastatic}) and (\ref{eqdstatic}) as $\alpha_0$ and  $\delta_0$.
Note also that  $\beta=-\beta_0$, $\gamma=-\gamma_0$ satisfy (\ref{QDstatic}) as well
as $\beta=\beta_0$, $\gamma=\gamma_0$ and that this change does not affect values
of $\alpha_0$ and $\delta_0$. We thus arrive at four distinct WANDs 
$(\alpha_0, \beta_0, \gamma_0, \pm \delta_0, 0)$,
$(\alpha_0, -\beta_0, -\gamma_0, \pm \delta_0, 0)$. The static black ring is thus of principal type I.
Furthermore, if we choose a frame with $\bl=(\alpha_0, \beta_0, \gamma_0,  \delta_0, 0)$ and
$\bn=(\alpha_0, \beta_0, \gamma_0,  -\delta_0, 0)$, we can see that all components with boost weight
2 and -2 vanish and thus the type is $(1,1)={\rm I}_i$.

\subsubsection{The stationary case}

Linear combination
\BE
(x-1) (x+1) (-1+\nu x) I_{t \psi y \phi} - \frac{1}{\lambda} (y-1) (1+\lambda) (-1+\nu y) I_{t \psi x \phi}=0
\EE
leads to
\BEA
\alpha\,\delta\, \bigl\{\!\!\!\! &&\!\!\!\! 
\alpha\,\gamma\,\sqrt {\lambda\,\nu}
 \left( x-y \right) ^{2} \left( -1+\lambda\,x \right)
+\epsilon\,\lambda\,{\it R}\, \left( y^2-1 \right) 
  \left[ \beta\,(1-\lambda\,y)+\gamma\,(\lambda\,x-1) \right] \nonumber \\ 
&+& \!\!\!\! \epsilon\,\nu\,\lambda\,{\it R}\,
 \left( 1+y \right)  \left[ \beta\,y \left( y-1 \right)  \left( -1+
\lambda\,y \right) +\gamma\, \left( -2\,xy+{x}^{2}+y \right)  \left( -
1+\lambda\,x \right)  \right]   \bigr\} = 0 .  \label{rcea}
\EEA
Assuming $\alpha\not= 0$ and $\delta  \not=0$, (\ref{rcea})  is a linear equation for $\alpha$  and by substituting the result in
$I_{xy \phi \psi}=0$ we obtain the quadratic equation for $\beta$ and  $\gamma$
\BEA
 \delta \epsilon \Bigl\{\!\!\! &&\!\!\! -{\beta}^{2}\lambda\, \left( y^2-1 \right) \left( -1+\lambda\,y \right)  
\left( -1+\nu\,y \right)  \left( -1+\lambda\,x \right) 
-{\gamma}^{2}\lambda\, \left( x^2-1 \right)  
 \left( -1+\lambda\,x \right) ^{2} \left( -1+\nu\,x \right) \nonumber \\  
&&\!\!\!
+\beta \gamma\, \Bigl[ 2(\nu-\lambda)(x-y)^2+\lambda^2(1-\lambda\,y)[x^2(x+y)+x-y]+2\lambda\,(1-\lambda\,x)^2
+2\lambda\,xy(-1+\lambda\,y)\label{rceQD} \\
&&\!\!\!
+2\lambda^2\nu\,xy(1-\lambda\,x)+\lambda\,\nu\,(1-\lambda\,y)[3x^2y(1-\lambda\,x)
+x^3(-1+\lambda\,y)-3x+y]+2\,\lambda\,\nu\,(x-y)\Bigr] \Bigr\} = 0. \nonumber
%
\EEA
Now the equation $\ell^a \ell_a = 0$ is linear in $\delta^2$ and using (\ref{rcea}) we can express
$\delta^2$ in terms of $\beta, \gamma, \epsilon$. Substituting this expression in
$I_{x \phi y \psi}=0$ gives
\BEA
&&\bigl[\gamma\, \left( -\lambda\,x+1 \right) -\beta\, \left( 1-\lambda\,y
 \right) \bigr]
  \Bigl\{ {\gamma}^{3}\nu\,
 \left( -1+\lambda\,x \right)  \left( -1+\lambda\,y \right) ^{2}
 \left( x-y \right) ^{2} 
-{\epsilon}^{2}\beta\,\lambda
\, \left( y^2-1 \right) ^{3}  \left( -1+\nu\,y
 \right) ^{3} \left( -1+\lambda\,y \right)
\nonumber \\
&&\ \ \ \ \ \ \ \ \ \ \ + {\epsilon}^{2}\gamma
\left( -1+\lambda\,x \right)  \left( y^2-1 \right) ^{2} \left( -1+\nu\,y \right) ^{2}
\left[ \nu\,x(1-\lambda\,y)(x-2y)+\lambda\,(1-\nu\,y)+y^2(\nu-\lambda)
\right]
%
%
  \Bigr\}
=0, \label{rcee2}
\EEA
which is linear in $\epsilon^2$. Solving this equation for $\epsilon^2$
and substituting to $\ell_a \ell^a=0$ leads to
\BEA
\!\!\!\!&&\!\!\!\! \delta^2 =   \Bigl\{ {\beta}^{3}\lambda\, \left( -1+\lambda\,x \right)  \left( y^2-1 \right) 
 \left( -1+\lambda\,y \right)  \left( -1+\nu\,y \right)
+\beta\, {\gamma}^{2}\lambda\, \left( -1+\lambda\,x \right) ^{2} 
\left( x^2-1 \right)  \left( -1+\nu\,x \right) \nonumber \\
\!\!\!\!&&\!\!\!\! -\gamma\, {\beta}^{2}\left( -1+\lambda\,x \right) 
\Bigl[ \lambda\nu\,(1-xy)[y(1-\lambda\,x)+x(1-\lambda\,y)]-\nu\,(x-y)^2+\lambda\,(1-\lambda\,y)(x^2-1)+\lambda\,(1-\lambda\,x)(y^2-1)\Bigr]
%
  \Bigr\} \nonumber \\
\!\!\!\!&&\!\!\!\!
/\Bigl\{  \left( -1+\nu\,x \right) ^{2} \left( x^2-1 \right) ^{2}
\Bigl[ \gamma \left( -1+\lambda\,x \right) 
\left[ \nu\,(x-y)^2(-1+\lambda\,y)+\lambda\,(y^2-1)(1-\nu\,y)\right]
%
%
 \nonumber \\ \!\!\!\!&&\ \ \ \ \ \ \ \ \ \ \ \ \ \ \  \ \ \ \ \ \ \ \ \ \ \ \ \ 
+\beta \lambda\, \left( y^2-1 \right)   \left( -1+\lambda\,y
 \right)  \left( -1+\nu\,y \right) \Bigr]  \Bigr\}. \label{rced2}
\EEA
Now we can explicitly express aligned null directions. Since if $\ell$ is aligned, then
also any multiple of $\ell$ is aligned, we can set $\beta=1$ without loss of generality.
Then we solve the quadratic equation (\ref{rceQD}) for $\gamma$ and express $\delta$ and $\epsilon$ from (\ref{rced2}) 
and  (\ref{rcee2}) and finally $\alpha$ from (\ref{rcea}).


Since the first condition in (\ref{HDnecessity}) is equivalence we have thus obtained expressions
for WANDs. In the static limit $\nu\rightarrow 0$ the solution corresponds to (\ref{eqastatic})--(\ref{QDstatic}).
Note, however, that for some values of $\lambda, \nu, x,y$ (see Figure 2) 
the value of $\epsilon ^2$ as determined by (\ref{rcee2}) is negative and thus
(\ref{rcea})--(\ref{rced2}) do not lead to  real WANDs at these points. Thus either spacetime is of type G there or there exists another 
solution of (\ref{BRI}) at these points.
This indicates that
the corresponding region in the vicinity of the axis $x=0$ is physically distinct from the rest.

\begin{figure}
\label{curve}
\begin{center}
\includegraphics[scale=0.5]{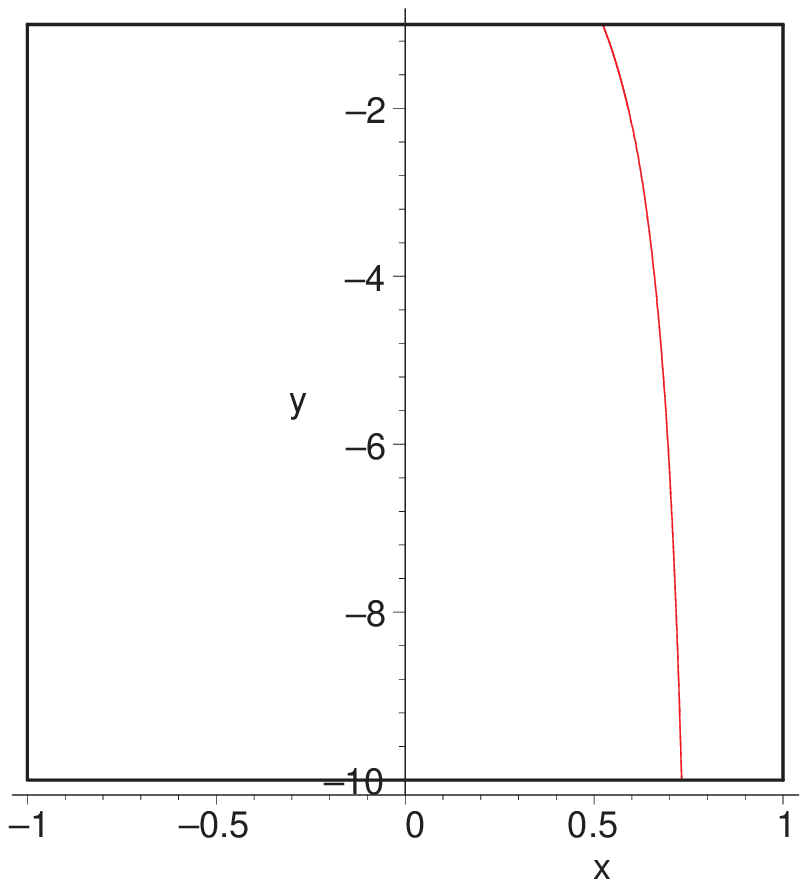}
\hspace{2cm}
\includegraphics[scale=0.5]{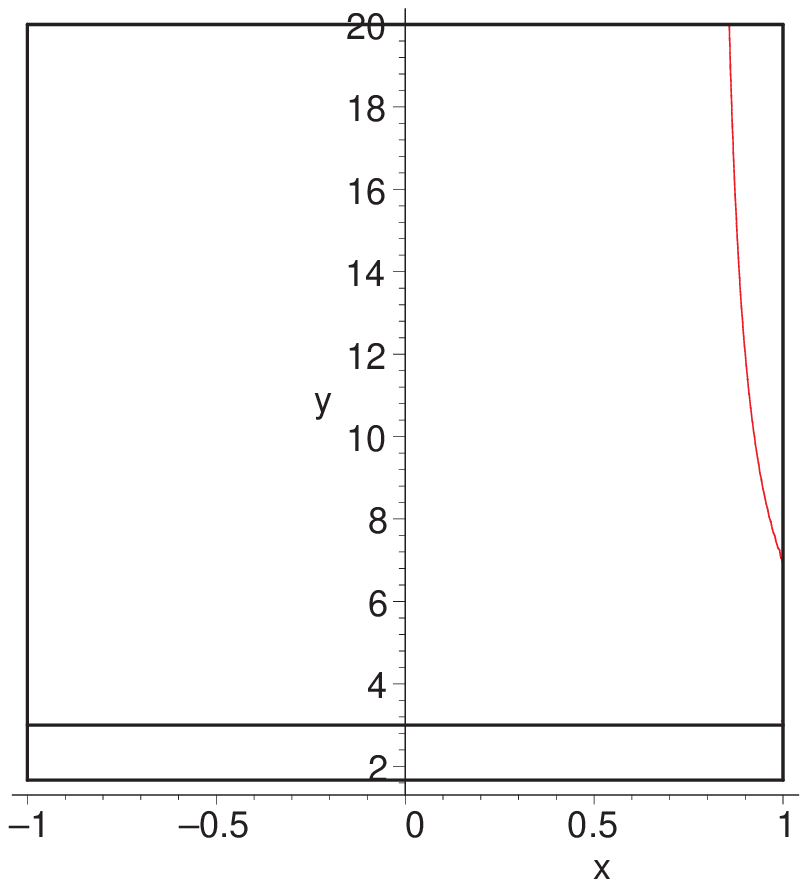}
\end{center}
\caption{A numerical example for $\lambda\,=3/5$, $\nu=1/3$: 
solutions of equations (\ref{rcea})--(\ref{rced2}) are real everywhere in the region ${\cal A}_3$ and only to the left from the indicated curve
in regions ${\cal A}_1$, ${\cal A}_2$.}
\label{diagram}
\end{figure}



\end{document}